\documentclass{article}
\usepackage[T1]{fontenc}

\usepackage[expansion=false]{microtype}
\usepackage{fontawesome5}
\usepackage[htt]{hyphenat}
\usepackage{graphicx}
\usepackage{float}
\usepackage{subcaption}
\usepackage{booktabs}
\usepackage{hyperref}

\PassOptionsToPackage{table}{xcolor}
\usepackage[preprint]{icml2027}

\usepackage{amsmath}
\usepackage{amssymb}
\usepackage{amsfonts}
\usepackage{mathtools}
\usepackage{bm}
\usepackage{amsmath,amsfonts,bm}

\def\eqref#1{equation~\ref{#1}}

\def\1{\bm{1}}

\DeclareMathAlphabet{\mathsfit}{\encodingdefault}{\sfdefault}{m}{sl}
\SetMathAlphabet{\mathsfit}{bold}{\encodingdefault}{\sfdefault}{bx}{n}

\usepackage{url}
\usepackage{multirow}
\usepackage{makecell}
\usepackage{nicefrac}
\usepackage{colortbl}
\usepackage{xspace}
\usepackage{enumitem}

\usepackage{tikz}
\usepackage{xcolor}
\usetikzlibrary{decorations.pathreplacing,calc,arrows.meta}

\definecolor{flat0}{RGB}{49,105,160}
\definecolor{flat1}{RGB}{103,151,194}
\definecolor{flat2}{RGB}{170,198,222}
\definecolor{flat3}{RGB}{218,231,241}

\definecolor{res0}{RGB}{75,135,105}

\definecolor{res1}{RGB}{105,175,125}
\definecolor{res2}{RGB}{190,225,195}
\definecolor{res3}{RGB}{235,246,236}
\definecolor{inter0}{RGB}{195,105,55}
\definecolor{inter1}{RGB}{230,135,65}
\definecolor{inter2}{RGB}{250,200,145}
\definecolor{inter3}{RGB}{253,238,220}

\definecolor{masked}{RGB}{245,245,245}
\newcommand{\mymodel}{\textsc{SimPos}}

\icmltitlerunning{Does Positional Encoding Matter for Semantic ID-based Generative Recommendation?}

\begin{document}

\twocolumn[
  \icmltitle{Does Positional Encoding Matter for Semantic ID-based\\Generative Recommendation?}

  \icmlsetsymbol{prior}{$\dagger$}

  \begin{icmlauthorlist}
    \icmlauthor{Yuqing Liu}{uic}
    \icmlauthor{Huiyuan Chen}{amzn,prior}
    \icmlauthor{Jingru Cheng}{stanford}
    \icmlauthor{Geo Zhang}{stanford}
    \icmlauthor{Philip S. Yu}{uic}
    \par\vspace{4pt}
    {\normalsize\textsuperscript{1}University of Illinois Chicago \quad \textsuperscript{2}Amazon \quad \textsuperscript{3}Stanford University}
    \par\vspace{2pt}
    {\small\texttt{\{yliu363, psyu\}@uic.edu}\quad\texttt{huiyuach@amazon.com}\quad\texttt{\{cjr63, gmzhang\}@stanford.edu}}
  \end{icmlauthorlist}

  \icmlkeywords{Generative Recommendation, Semantic IDs, Positional Encoding}

  \vskip 0.3in
]

\printAffiliationsAndNotice{\textsuperscript{$\dagger$}Work done prior to joining Amazon. }

\begin{abstract}
Semantic IDs have emerged as an effective representation for generative recommendation, where each item is represented by multiple discrete codes.
These codes are typically serialized along the user interaction sequence into a flattened token sequence for generative modeling.
However, such a formulation couples two structurally different sources of positional
variation: the sequential order of user interactions and the hierarchical
order induced by residual quantization.
We argue that these two structures should be modeled on their respective
positional axes.
Based on this insight, we propose \mymodel, a simple structure-aware positional modeling method that replaces the Transformer backbone's positional encoding with separate interaction-level and residual-level attention biases, requiring no additional learnable parameters.
Experiments on four real-world recommendation datasets with different
backbones and residual tokenizers show that \mymodel\
consistently outperforms existing positional encoding methods,
with relative gains of up to $+24.88\%$ in Recall@5.
These results indicate that positional encoding matters in semantic ID-based generative recommendation, and that directly inheriting it from the Transformer backbone is suboptimal.
Our code and data are available at \href{https://github.com/KevinC2015/SimPos}{\textcolor{blue!60!black}{\faGithub\ \texttt{github.com/KevinC2015/SimPos}}}.
\end{abstract}

\section{Introduction}

\begin{figure}[t]
    \centering
    \includegraphics[width=\columnwidth]{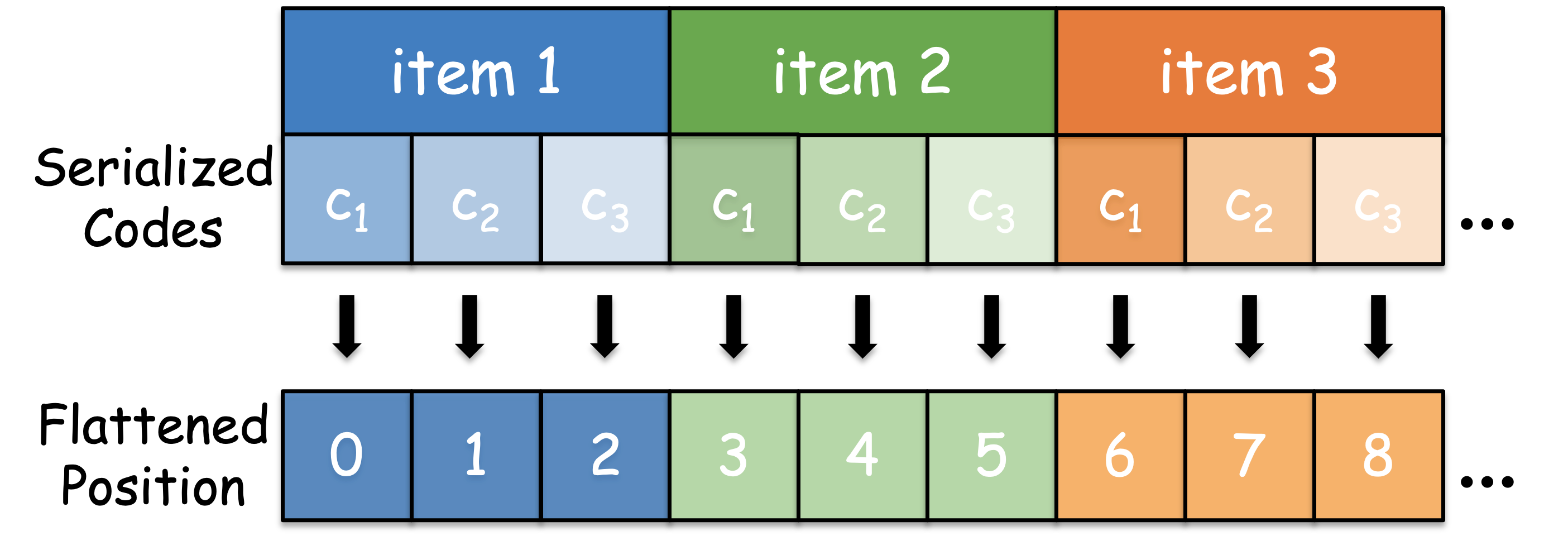}
    \caption{Flattened positional indexing for multi-code semantic IDs. Each code has an interaction position and a residual position, but receives a single index along the flattened sequence.}
    \label{fig:flatten}
\end{figure}
Recent advances in generative recommendation formulate next-item recommendation as a sequence generation problem~\citep{p5,hou2024large,hstu,bao2023tallrec,wang2024eager}.
A key challenge in this paradigm is how to represent discrete items in a form that is both semantically meaningful and suitable for generation. Semantic IDs (SIDs) have emerged as a promising solution by representing each item with a short sequence of discrete codes derived from its semantic representation~\citep{tiger,rpg,diger,liu2026rethinking}.
A typical SID-based generative recommender first encodes item content into continuous semantic embeddings, quantizes them into discrete semantic IDs, and serializes the resulting codes into the user interaction sequence for generative modeling.
Among different tokenization schemes, residual quantization has been widely adopted~\citep{rqvae,ju2025generative,zheng2024adapting,card}, constructing semantic IDs through successive residual refinement.

Despite substantial progress in learning effective semantic IDs, how these multi-code item representations should be positionally modeled in user interaction sequences has received limited attention.
As illustrated in Figure~\ref{fig:flatten}, the residual codes of interacted items are serialized into a single token sequence, with each code assigned an index along the flattened sequence.
This serialization, however, maps two distinct positional structures onto a single token-level positional axis.
The first is the \textbf{\emph{interaction position}}, which specifies where an item occurs in the interaction sequence, capturing the sequential order across items.
The second is the \textbf{\emph{residual position}}, which indicates the quantization level of a code, reflecting coarse-to-fine semantic refinement.
Consequently, while each semantic code carries both an interaction-level and a residual-level coordinate, conventional positional modeling treats positions along the flattened sequence uniformly, without explicitly distinguishing these two structural roles.

This distinction becomes particularly important for relative positions between semantic codes.
On the one hand, codes from the same pair of adjacent items can be one to several positions apart in the flattened sequence, depending on their residual levels.
On the other hand, the last code of one item and the first code of the next are one position apart, just like two adjacent codes within the same item.
A single token-level offset, therefore, does not by itself tell these structurally different relations apart.
These observations suggest that interaction order and residual order should both be preserved, but modeled on separate positional axes rather than through the flattened relative displacement alone.

Motivated by this insight, we propose \mymodel, a simple structure-aware positional modeling method for semantic ID-based generative recommendation.
Rather than relying on offsets within the flattened one-dimensional sequence, \mymodel\ replaces the backbone's positional encoding with two additive attention biases.
They penalize the distance between items in the interaction sequence and the distance between residual levels, respectively.
Moreover, \mymodel\ introduces no additional learnable parameters and can be plugged into both encoder--decoder and decoder-only Transformers without other architectural changes.
Experiments on four datasets with two backbones and two residual tokenizers show that \mymodel\ consistently outperforms existing positional encoding methods, with relative gains of up to $+24.88\%$ in Recall@5.
These results indicate that positional encoding is an important yet underexplored factor in semantic ID-based generative recommendation.
Our main contributions are summarized as follows:
\begin{itemize}[leftmargin=*, topsep=2pt, itemsep=2pt, parsep=0pt, partopsep=0pt]
    \item We identify a structural mismatch between flattened positional encodings and residual semantic IDs. Conventional positional encodings operate on a single flattened token-level axis without explicitly modeling the distinct structural roles of interaction order and residual order.
    \item We propose \mymodel, a plug-and-play structure-aware positional modeling method that encodes interaction-level and residual-level distances as two separate additive attention biases, without additional learnable parameters.
    \item We conduct extensive experiments on four real-world datasets using different Transformer backbones and residual tokenizers. \mymodel\ consistently outperforms existing positional encoding methods under diverse settings, demonstrating its effectiveness and general applicability.
\end{itemize}

\section{Related Work}

\subsection{Semantic ID-based Generative Recommendation}
Unlike traditional atomic ID-based recommendation~\citep{sasrec,bert4rec,he2017neural,chen2022denoising},
generative recommendation formulates recommendation as a sequence generation task~\citep{li2024large,hou2024large,hstu,li2023gpt4rec,wang2024eager}.
A representative framework is TIGER~\citep{tiger}, which represents each item as a sequence of discrete codes learned through residual quantization and autoregressively generates the SID of the next item.
Following this paradigm, subsequent studies have explored diverse SID construction methods and generative modeling strategies.
On the construction side, LETTER~\citep{letter} incorporates collaborative signals and code-assignment diversity into tokenization. DIGER~\citep{diger} introduces differentiable semantic indexing to better align tokenization with recommendation objectives.
On the generation side, recent methods move beyond autoregressive decoding through parallel decoding~\citep{rpg,liu2026rethinking} or diffusion-based generation~\citep{lladarec, liu2026diffgrm, mdgr}.
Beyond SID construction and generation, recent studies have also explored modeling SID structure. CoFiRec~\citep{wei2026cofirec} constructs coarse-to-fine SIDs from different semantic attributes and collaborative signals, incorporating learnable level and item-position embeddings. DiscRec~\citep{liu2025discrec} introduces item-level position embeddings to encode the position of each code within its semantic ID, together with localized intra-item attention in a separate collaborative branch.
BARGE~\citep{zeng2026bridging} addresses the item-boundary gap through Item Context-Aware Attention, which pools each item's code embeddings and injects the resulting context into its tokens through gated residual connections. This module operates at the embedding level and retains the backbone's positional encoding.
While these methods incorporate structural information through additional learnable embeddings or modules, we instead replace the backbone's positional encoding with parameter-free relative attention biases that explicitly model such structure.

\subsection{Positional Encoding in Transformers}
Positional encodings in Transformers can be broadly grouped into absolute and relative ones.
Absolute positional encodings assign each token a position-specific representation and inject it into the input embeddings~\citep{vaswani2017attention,devlin2019bert}.
In contrast, relative positional encodings model positional relationships between tokens and have been extensively studied in various forms~\citep{shaw2018self,transformerxl,t5,deberta,tupe}.
Specifically, \citet{vaswani2017attention} use a fixed, sinusoidal positional encoding composed of alternating sine and cosine functions of varying frequencies. \citet{devlin2019bert} utilize learned absolute positional embeddings, adding them directly to the token and segment embeddings before feeding into the Transformer layers. Moreover, \citet{shaw2018self} introduce learned representations of relative positions into self-attention.
RoPE~\citep{rope} encodes positional information through position-dependent rotations of query and key representations, while
ALiBi~\citep{alibi} directly introduces distance-dependent biases into attention scores.
These methods define positions along a single token sequence, and how to model positional information for the multi-code item structure in SID-based generative recommendation remains underexplored.
Our work addresses this gap by explicitly modeling both the interaction order across items and the residual order within each item.

\section{Methodology}
\label{sec:methodology}

Let $\mathcal{U}$ and $\mathcal{I}$ denote the user set and item set, respectively.
Given a user's chronologically ordered interaction sequence
$\mathcal{S}_u = (i_1, i_2, \dots, i_T)$, where $i_t \in \mathcal{I}$
denotes the $t$-th interacted item, our goal is to predict the next item $i_{T+1}$.
Following prior work on semantic ID-based generative recommendation~\citep{tiger,onerec}, we represent each item as a sequence of discrete semantic codes and formulate next-item prediction as autoregressive generation of the target item's semantic ID.

\subsection{Preliminaries}
\label{sec:prelim}

\paragraph{Semantic ID Construction.}
We focus on residual-based semantic tokenization (e.g., RQ-VAE~\citep{rqvae}, RQ-KMeans~\citep{qarm}), which represents each item with an ordered sequence of discrete codes through residual vector quantization.
Given an item $i \in \mathcal{I}$, we first obtain its dense semantic embedding $\mathbf{x}_i \in \mathbb{R}^d$ from its content information (e.g., title, category, description).
A residual quantizer then hierarchically tokenizes the item embedding using $K$ codebooks, each containing $M$ codewords.
The first codebook quantizes the original embedding, while each subsequent codebook quantizes the residual left by the previous levels. The resulting semantic ID is $\mathbf{c}_i = (c_{i,1}, c_{i,2}, \ldots, c_{i,K})$, where $c_{i,k}\in \{1,\ldots,M\}$ denotes the selected index in the $k$-th codebook.

\paragraph{Interaction Sequence Representation.}
Given the interaction sequence $\mathcal{S}_u=(i_1,\ldots,i_T)$, we denote the $k$-th code of the $t$-th interacted item $i_t$ as $c_{t,k}$ for simplicity. The semantic IDs of the historical items are then serialized into a single flattened token sequence:
\begin{equation}
\begin{split}
\mathcal{C}_u = (&c_{1,1},\ldots,c_{1,K},\,
 c_{2,1},\ldots,c_{2,K},\\
 &\ldots,\,
 c_{T,1},\ldots,c_{T,K}).
\end{split}
\label{eq:serialization}
\end{equation}
The resulting $TK$ semantic codes then serve as the input to the Transformer model.

\subsection{Rethinking Positional Structure in Semantic IDs}
\label{sec:rethinking_position}
The serialization in Eq.~(\ref{eq:serialization}) maps each code $c_{t,k}$, indexed by its interaction position $t$ and residual position $k$, onto a single position along the flattened sequence, given by $p(t,k) = (t-1)K + k$.
{For a query--key pair $(c_{t,k}, c_{s,j})$, we define their flattened relative position as}
\begin{equation}
r(c_{t,k}, c_{s,j}) = p(t,k)-p(s,j)=(t-s)K+(k-j),
\label{eq:flattened_relative_position}
\end{equation}
where $t-s$ and $k-j$ are their relative differences in interaction
and residual positions, respectively.

\begin{figure*}[t]
\centering

\resizebox{\textwidth}{!}{%
\begin{tikzpicture}[
    x=0.38cm,
    y=0.38cm,
    >=Stealth
]

\begin{scope}[shift={(0,0)}]

\foreach \q in {0,...,8}{
    \foreach \k in {0,...,8}{

        \ifnum\k>\q
            \fill[masked] (\k,-\q) rectangle ++(1,-1);

        \else
            \pgfmathtruncatemacro{\dist}{\q-\k}

            \ifnum\dist=0
                \fill[flat0] (\k,-\q) rectangle ++(1,-1);
            \else
                \ifnum\dist=1
                    \fill[flat1] (\k,-\q) rectangle ++(1,-1);
                \else
                    \ifnum\dist=2
                        \fill[flat2] (\k,-\q) rectangle ++(1,-1);
                    \else
                        \fill[flat3] (\k,-\q) rectangle ++(1,-1);
                    \fi
                \fi
            \fi

            \node[font=\scriptsize]
                at (\k+0.5,-\q-0.5)
                {\dist};
        \fi

        \draw[white,line width=0.25pt]
            (\k,-\q) rectangle ++(1,-1);
    }
}

\draw[line width=0.65pt]
    (0,0) rectangle (9,-9);

\draw[line width=1.15pt] (3,0) -- (3,-9);
\draw[line width=1.15pt] (6,0) -- (6,-9);
\draw[line width=1.15pt] (0,-3) -- (9,-3);
\draw[line width=1.15pt] (0,-6) -- (9,-6);

\node[font=\normalsize] at (1.5,0.65) {$i_1$};
\node[font=\normalsize] at (4.5,0.65) {$i_2$};
\node[font=\normalsize] at (7.5,0.65) {$i_3$};

\node[rotate=90,font=\normalsize] at (-0.65,-1.5) {$i_1$};
\node[rotate=90,font=\normalsize] at (-0.65,-4.5) {$i_2$};
\node[rotate=90,font=\normalsize] at (-0.65,-7.5) {$i_3$};

\node[
    font=\small\bfseries,
    text width=5.0cm,
    align=center
]
at (4.5,-10.1)
{(a) Flattened Token Position};

\end{scope}

\begin{scope}[shift={(26,0)}]

\foreach \q in {0,...,8}{
\foreach \k in {0,...,8}{

    \pgfmathtruncatemacro{\qres}{mod(\q,3)}
    \pgfmathtruncatemacro{\kres}{mod(\k,3)}

    \fill[masked]
        (\k,-\q) rectangle ++(1,-1);

    \ifnum\k>\q
    \else

        \pgfmathtruncatemacro{\dist}{abs(\qres-\kres)}

        \ifnum\dist=0
            \fill[res0]
                (\k,-\q) rectangle ++(1,-1);
        \else
            \ifnum\dist=1
                \fill[res1]
                    (\k,-\q) rectangle ++(1,-1);
            \else
                \fill[res2]
                    (\k,-\q) rectangle ++(1,-1);
            \fi
        \fi

        \node[font=\scriptsize]
            at (\k+0.5,-\q-0.5)
            {\dist};

    \fi

    \draw[white,line width=0.25pt]
        (\k,-\q) rectangle ++(1,-1);
}
}

\draw[line width=0.65pt]
    (0,0) rectangle (9,-9);

\draw[line width=1.15pt] (3,0) -- (3,-9);
\draw[line width=1.15pt] (6,0) -- (6,-9);
\draw[line width=1.15pt] (0,-3) -- (9,-3);
\draw[line width=1.15pt] (0,-6) -- (9,-6);

\node[font=\normalsize] at (1.5,0.65) {$i_1$};
\node[font=\normalsize] at (4.5,0.65) {$i_2$};
\node[font=\normalsize] at (7.5,0.65) {$i_3$};

\node[rotate=90,font=\normalsize] at (-0.65,-1.5) {$i_1$};
\node[rotate=90,font=\normalsize] at (-0.65,-4.5) {$i_2$};
\node[rotate=90,font=\normalsize] at (-0.65,-7.5) {$i_3$};

\node[
    font=\small\bfseries,
    text width=5.0cm,
    align=center
]
at (4.5,-10.1)
{(c) Residual-level Position};

\end{scope}

\begin{scope}[shift={(13,0)}]

\foreach \q in {0,...,8}{
    \foreach \k in {0,...,8}{

        \pgfmathtruncatemacro{\qitem}{floor(\q/3)}
        \pgfmathtruncatemacro{\kitem}{floor(\k/3)}

        \ifnum\k>\q

            \fill[masked]
                (\k,-\q) rectangle ++(1,-1);

        \else

            \pgfmathtruncatemacro{\dist}{\qitem-\kitem}

            \ifnum\dist=0
                \fill[inter0]
                    (\k,-\q) rectangle ++(1,-1);
            \else
                \ifnum\dist=1
                    \fill[inter1]
                        (\k,-\q) rectangle ++(1,-1);
                \else
                    \fill[inter2]
                        (\k,-\q) rectangle ++(1,-1);
                \fi
            \fi

            \node[font=\scriptsize]
                at (\k+0.5,-\q-0.5)
                {\dist};

        \fi

        \draw[white,line width=0.25pt]
            (\k,-\q) rectangle ++(1,-1);
    }
}

\draw[line width=0.65pt]
    (0,0) rectangle (9,-9);

\draw[line width=1.15pt] (3,0) -- (3,-9);
\draw[line width=1.15pt] (6,0) -- (6,-9);
\draw[line width=1.15pt] (0,-3) -- (9,-3);
\draw[line width=1.15pt] (0,-6) -- (9,-6);

\node[font=\normalsize] at (1.5,0.65) {$i_1$};
\node[font=\normalsize] at (4.5,0.65) {$i_2$};
\node[font=\normalsize] at (7.5,0.65) {$i_3$};

\node[rotate=90,font=\normalsize] at (-0.65,-1.5) {$i_1$};
\node[rotate=90,font=\normalsize] at (-0.65,-4.5) {$i_2$};
\node[rotate=90,font=\normalsize] at (-0.65,-7.5) {$i_3$};

\node[
    font=\small\bfseries,
    text width=5.0cm,
    align=center
]
at (4.5,-10.1)
{(b) Interaction-level Position};

\end{scope}

\node[
    font=\large\bfseries,
    align=center
]
at (4.5,2.25)
{Relative Position};

\node[
    font=\large\bfseries,
    align=center
]
at (24.0,2.25)
{\mymodel};

\end{tikzpicture}
}

\caption{
Illustration of relative positional encoding and our proposed
\mymodel\ for semantic ID-based generative recommendation (e.g., a sequence of 3 items  $[i_1, i_2, i_3]$, with each represented by 3 codes).
(a) Relative positions are computed along the flattened sequence.
\mymodel\ explicitly distinguishes two positional structures:
(b) interaction-level positions, which capture relative differences across items in the interaction sequence, and
(c) residual-level positions, which capture relative differences across residual levels. Only the lower triangle (causal attention) is shown for clarity.
}
\label{fig:simpos}
\vspace{-4pt}
\end{figure*}
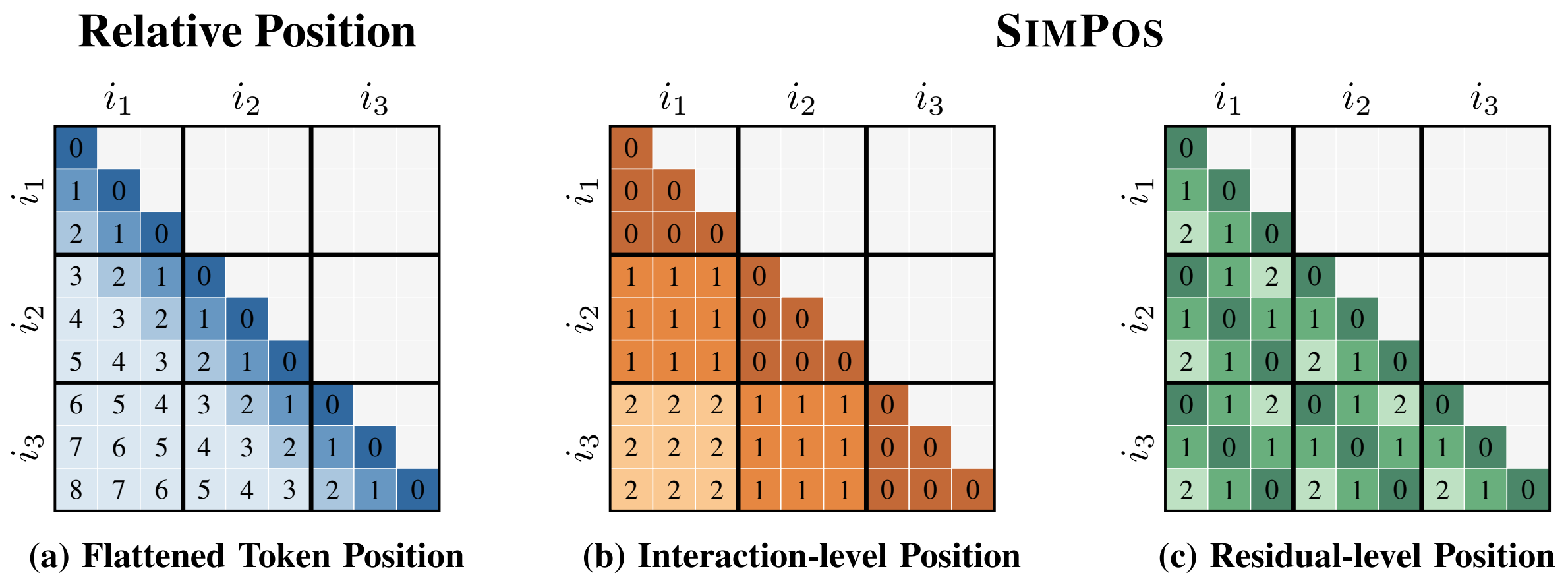

Eq.~(\ref{eq:flattened_relative_position}) shows that flattening collapses two structurally different sources of positional variation into a single scalar.
As illustrated in Figure~\ref{fig:simpos}(a), this leads to {three} observations:

\emph{(1) Different structural relations are represented on the same positional scale.}
With $K=3$, two codes at the same residual level from adjacent items
have $r=3$ ($t-s=1, k=j$), whereas two adjacent residual codes within the same item have $r=1$ ($t=s, k-j=1$).
{Consequently, one step in interaction position spans $K$ token positions, whereas one step in residual position spans only one.}

\emph{(2) One interaction-level relation maps to multiple flattened positions.}
For a pair of adjacent items, the interaction-position difference is fixed at $t-s=1$.
Their codes, however, yield values of $r$ ranging from $1$ to $5$, depending on the residual levels being compared.

\emph{(3) Different structural relations map to the same flattened position.}
For example, $r(c_{1,2}, c_{1,1}) = r(c_{2,1}, c_{1,3}) = 1$, although the former pair lies within one item and the latter spans two adjacent items.
The flattened relative displacement alone cannot distinguish these two relations.

These observations highlight that both interaction order and residual order carry meaningful positional information.
The former captures the sequential order of user interactions, while the latter reflects the hierarchical refinement order induced by residual quantization.
Although the flattened absolute position uniquely determines both coordinates, their relative differences are combined on a single flattened positional axis.
This motivates a structure-aware positional formulation that explicitly models interaction-level and residual-level relations separately, providing a structural inductive bias tailored to SID sequences.

\subsection{SimPos: Structure-Aware Positional Encoding}
\label{sec:simpos}

Based on the above observations, we propose \textbf{\mymodel}, a
structure-aware positional encoding approach that decouples interaction-level and residual-level positional relations in the sequences of semantic IDs.

\paragraph{Interaction-level Position.}
We define the interaction-level positional bias of $(c_{t,k}, c_{s,j})$ as
\begin{equation}
B_{\text{int}}(c_{t,k}, c_{s,j})
= -\lambda_{\text{int}} \cdot |t-s|,
\label{eq:simpos_interaction}
\end{equation}
where $\lambda_{\text{int}}\geq 0$ controls the strength of the interaction position signal.
As shown in Figure~\ref{fig:simpos}(b), all code pairs between the same two items receive the same interaction bias, regardless of residual levels.
{This bias encourages each code to attend more to codes from recent items, reflecting the recency prior commonly used in sequential recommendation~\citep{liu2018stamp,wu2019session}.}

\paragraph{Residual-level Position.}
Similarly, the residual-level positional bias of $(c_{t,k}, c_{s,j})$ is defined as
\begin{equation}
    {B}_{\text{res}}(c_{t,k}, c_{s,j}) = -\lambda_{\text{res}} \cdot |k-j|,
    \label{eq:simpos_residual}
\end{equation}
where $\lambda_{\text{res}}\geq 0$ controls the strength of the residual position signal.
As illustrated in Figure~\ref{fig:simpos}(c), the bias depends only on the distance between residual levels, independent of their interaction positions.
This bias encourages each code to attend more to codes at nearby residual levels, which capture semantics at similar granularity.

\paragraph{Unified Attention Formulation.} When using \mymodel, we do not add position embeddings in the neural networks. The only
modification we apply is after the query--key dot product.
Combining the two positional components, the attention logit for the
query--key pair $(c_{t,k}, c_{s,j})$ is given by
\begin{equation}
\begin{split}
\ell(c_{t,k},c_{s,j})
={}&
\frac{\mathbf{q}_{t,k}^{\top}\mathbf{k}_{s,j}}{\sqrt{d_h}}
+
B_{\text{int}}(c_{t,k},c_{s,j})\\
&+
B_{\text{res}}(c_{t,k},c_{s,j}),
\end{split}
\label{eq:simpos_attention}
\end{equation}
where $\mathbf{q}_{t,k}$ and $\mathbf{k}_{s,j}$ denote the query and key representations, and $d_h$ is the attention head dimension.
By design, \mymodel\ introduces \textit{no additional learnable parameters}. The two coefficients are fixed and shared across all attention heads and layers.
As \mymodel\ operates directly on attention logits, it can be readily integrated into different Transformer architectures without modifying other model components or the training objective.
In this way, \mymodel\ measures interaction-level and residual-level distances on separate axes, each with its own scale.
The bias between two codes is thus determined by their structural relation rather than their offset in the flattened sequence.

\section{Experiments}
\label{sec:experiment}

In this section, we conduct experiments to answer the following research questions:
\textbf{RQ1}: Does \mymodel\ outperform existing positional encodings across different datasets, Transformer backbones, and residual tokenizers?
\textbf{RQ2}: How do the interaction-level and residual-level components each contribute to the performance of \mymodel?
\textbf{RQ3}: How sensitive is \mymodel\ to the strengths of the two positional biases?
\textbf{RQ4}: Does \mymodel\ remain effective with different semantic embedding models?
\textbf{RQ5}: What are the benefits of the two-axis relative bias in \mymodel?

\subsection{Experimental Setup}
\label{sec:experiment_setup}
\textbf{Datasets.}
We evaluate \mymodel\ on four categories from the Amazon Reviews 2023\footnote{https://amazon-reviews-2023.github.io/}~\citep{hou2026bridging}: Beauty, Toys, Sports, and CDs.
Preprocessing details and dataset statistics are provided in Appendix~\ref{app:datasets}.

\textbf{Backbones and Tokenizers.}
To examine the general applicability of \mymodel{}, we consider two Transformer backbones and two residual tokenizers.
For the encoder--decoder backbone, we adopt T5~\citep{t5} following TIGER~\citep{tiger}, whose original positional encoding is the bucketed relative position bias. For the decoder-only backbone, we adopt GPT-2~\citep{radford2019language}, whose original positional encoding is the learnable absolute position embedding.
For semantic ID construction, we use RQ-VAE~\citep{rqvae} and RQ-KMeans~\citep{qarm}.
In all settings, we concatenate the title, description, and category of each item and encode them with \texttt{Qwen3-Embedding-0.6B}\footnote{\label{fn:qwen}\url{https://huggingface.co/Qwen/Qwen3-Embedding-0.6B}} into a 1024-dimensional embedding.
The residual tokenizer then quantizes each embedding with $K=4$ codebooks, each containing $M=256$ codewords, so that every item is represented by a semantic ID of four codes.
Following TIGER~\citep{tiger}, the last code is a deduplication code that distinguishes items sharing the same first three codes.

\textbf{Baselines.}
Since \mymodel\ only modifies how positional information is modeled, we compare it with representative positional encodings (PEs). Within each setting, all methods share the same semantic IDs and training configuration. Specifically, we consider
\textbf{(1) LPE}, learnable absolute position embeddings added to the input embeddings, the default in GPT-2~\citep{radford2019language};
\textbf{(2) RelPE}, bucketed relative position biases added to attention logits, the default in T5~\citep{t5};
\textbf{(3) RoPE}~\citep{rope}, which rotates query and key representations according to their positions; and
\textbf{(4) ALiBi}~\citep{alibi}, which adds linear distance-based biases to attention logits.
All baselines compute positions on the flattened sequence.

\textbf{Evaluation.}
We adopt the leave-last-out evaluation protocol~\citep{sasrec,p5,zheng2024adapting}.
For each user, the last item is used for testing, the second-to-last item for validation, and the remaining interactions for training.
We report Recall@5 (R@5) and NDCG@5 (N@5).
Each model is trained with five independent seeds, and we assess statistical significance using one-sided paired t-tests, pairing \mymodel\ with the best baseline by seed.
We apply Holm correction to the p-values.

\textbf{Implementation Details.}
For the backbones, we use a $4$-layer, $6$-head T5 encoder--decoder and a $4$-layer, $4$-head GPT-2 decoder. Both architectures share a hidden dimension of $128$, and we keep the most recent $50$ interactions of each user.
All models are trained with AdamW at a learning rate of $3 \times 10^{-4}$ and a batch size of $512$ for up to $200$ epochs, with early stopping based on validation N@5.
For \mymodel, $\lambda_{\text{int}}$ and $\lambda_{\text{res}}$ are searched over $\{0.4, 0.6, 0.8\}$ and $\{0.1, 0.2, 0.3\}$, and the selected values are listed in Appendix~\ref{app:coefficients}.
During inference, we use beam search with a beam size of $30$.
All experiments are conducted on a single NVIDIA A100 (80GB) GPU.

\subsection{Overall Performance (RQ1)}
\label{sec:overall_performance}
\begin{table*}[t]
  \centering
  \caption{Performance of different positional encodings (PE) on four datasets with two backbones and two tokenizers. Best in \textbf{bold}, second best \underline{underlined}. \textit{Improv.} is the relative gain of \mymodel\ over the best baseline. Results are means over five training seeds (std $\leq 0.0008$). \mymodel\ achieves statistically significant improvements under a paired t-test (Holm-adjusted $p < 0.005$).}
  \vspace{-2pt}
  \label{tab:pe_comparison}
  \setlength{\tabcolsep}{1.5mm}
  \resizebox{\textwidth}{!}{
  \begin{tabular}{@{}llcccccccc@{}}
    \toprule
    \multirow{2.5}{*}{\textbf{Tokenizer}} & \multirow{2.5}{*}{\textbf{PE}} & \multicolumn{2}{c}{\textbf{Beauty}} & \multicolumn{2}{c}{\textbf{Toys}} & \multicolumn{2}{c}{\textbf{Sports}} & \multicolumn{2}{c}{\textbf{CDs}} \\
    \cmidrule(l){3-4} \cmidrule(l){5-6} \cmidrule(l){7-8} \cmidrule(l){9-10}
    & & R@5 & N@5 & R@5 & N@5 & R@5 & N@5 & R@5 & N@5 \\
    \midrule\midrule
    \multicolumn{10}{@{}c}{\textit{T5 backbone (Original positional encoding is RelPE)}} \\
    \midrule
    \multirow{7}{*}{RQ-VAE}
      & LPE & 0.0202 & 0.0129 & 0.0182 & 0.0109 & 0.0183 & 0.0122 & 0.0301 & 0.0183 \\
      & RelPE & 0.0211 & 0.0134 & \underline{0.0187} & \underline{0.0119} & 0.0197 & 0.0126 & 0.0303 & 0.0187 \\
      & RoPE & 0.0208 & 0.0132 & 0.0176 & 0.0103 & \underline{0.0199} & \underline{0.0130} & \underline{0.0311} & \underline{0.0193} \\
      & ALiBi & \underline{0.0213} & \underline{0.0136} & 0.0186 & 0.0111 & 0.0190 & 0.0117 & 0.0308 & 0.0190 \\
      & \textbf{\mymodel} & \textbf{0.0228} & \textbf{0.0152} & \textbf{0.0207} & \textbf{0.0135} & \textbf{0.0248} & \textbf{0.0154} & \textbf{0.0345} & \textbf{0.0214} \\
    \rowcolor{gray!12}
    \cellcolor{white}
      & \textit{Improv.} & \textit{+7.04\%} & \textit{+11.76\%} & \textit{+10.70\%} & \textit{+13.45\%} & \textit{+24.62\%} & \textit{+18.46\%} & \textit{+10.93\%} & \textit{+10.88\%} \\
    \cmidrule{1-10}
    \multirow{7}{*}{RQ-KMeans}
      & LPE & 0.0203 & 0.0126 & 0.0209 & 0.0133 & 0.0200 & 0.0130 & 0.0294 & 0.0182 \\
      & RelPE & 0.0214 & 0.0135 & \underline{0.0220} & \underline{0.0144} & 0.0212 & \underline{0.0138} & \underline{0.0311} & 0.0192 \\
      & RoPE & 0.0210 & 0.0138 & 0.0218 & 0.0139 & 0.0206 & 0.0136 & 0.0304 & \underline{0.0203} \\
      & ALiBi & \underline{0.0217} & \underline{0.0139} & 0.0211 & 0.0143 & \underline{0.0217} & \underline{0.0138} & 0.0301 & 0.0186 \\
      & \textbf{\mymodel} & \textbf{0.0271} & \textbf{0.0173} & \textbf{0.0231} & \textbf{0.0159} & \textbf{0.0261} & \textbf{0.0172} & \textbf{0.0331} & \textbf{0.0215} \\
    \rowcolor{gray!12}
    \cellcolor{white}
      & \textit{Improv.} & \textit{+24.88\%} & \textit{+24.46\%} & \textit{+5.00\%} & \textit{+10.42\%} & \textit{+20.28\%} & \textit{+24.64\%} & \textit{+6.43\%} & \textit{+5.91\%} \\
    \midrule
    \multicolumn{10}{@{}c}{\textit{GPT-2 backbone (Original positional encoding is LPE)}} \\
    \midrule
    \multirow{7}{*}{RQ-VAE}
      & LPE & 0.0197 & 0.0128 & \underline{0.0172} & \underline{0.0102} & \underline{0.0213} & \underline{0.0137} & \underline{0.0285} & 0.0177 \\
      & RelPE & 0.0199 & 0.0128 & 0.0163 & 0.0097 & 0.0211 & 0.0136 & 0.0276 & \underline{0.0190} \\
      & RoPE & \underline{0.0204} & 0.0123 & 0.0161 & 0.0101 & 0.0204 & 0.0120 & 0.0278 & 0.0171 \\
      & ALiBi & 0.0185 & \underline{0.0130} & 0.0165 & 0.0098 & 0.0208 & 0.0127 & 0.0264 & 0.0167 \\
      & \textbf{\mymodel} & \textbf{0.0230} & \textbf{0.0150} & \textbf{0.0185} & \textbf{0.0120} & \textbf{0.0234} & \textbf{0.0156} & \textbf{0.0319} & \textbf{0.0203} \\
    \rowcolor{gray!12}
    \cellcolor{white}
      & \textit{Improv.} & \textit{+12.75\%} & \textit{+15.38\%} & \textit{+7.56\%} & \textit{+17.65\%} & \textit{+9.86\%} & \textit{+13.87\%} & \textit{+11.93\%} & \textit{+6.84\%} \\
    \cmidrule{1-10}
    \multirow{7}{*}{RQ-KMeans}
      & LPE & \underline{0.0208} & 0.0133 & 0.0195 & \underline{0.0125} & \underline{0.0214} & 0.0146 & \underline{0.0303} & 0.0191 \\
      & RelPE & 0.0197 & 0.0131 & \underline{0.0200} & 0.0118 & \underline{0.0214} & \underline{0.0157} & 0.0290 & \underline{0.0201} \\
      & RoPE & 0.0207 & \underline{0.0139} & 0.0191 & 0.0124 & 0.0204 & 0.0152 & 0.0283 & 0.0191 \\
      & ALiBi & \underline{0.0208} & 0.0131 & 0.0190 & 0.0122 & 0.0200 & 0.0135 & 0.0288 & 0.0195 \\
      & \textbf{\mymodel} & \textbf{0.0227} & \textbf{0.0152} & \textbf{0.0219} & \textbf{0.0139} & \textbf{0.0254} & \textbf{0.0169} & \textbf{0.0343} & \textbf{0.0215} \\
    \rowcolor{gray!12}
    \cellcolor{white}
      & \textit{Improv.} & \textit{+9.13\%} & \textit{+9.35\%} & \textit{+9.50\%} & \textit{+11.20\%} & \textit{+18.69\%} & \textit{+7.64\%} & \textit{+13.20\%} & \textit{+6.97\%} \\
    \bottomrule
  \end{tabular}}
  \vspace{-4pt}
\end{table*}

\begin{figure*}[t]
  \centering
  \includegraphics[width=0.9\textwidth]{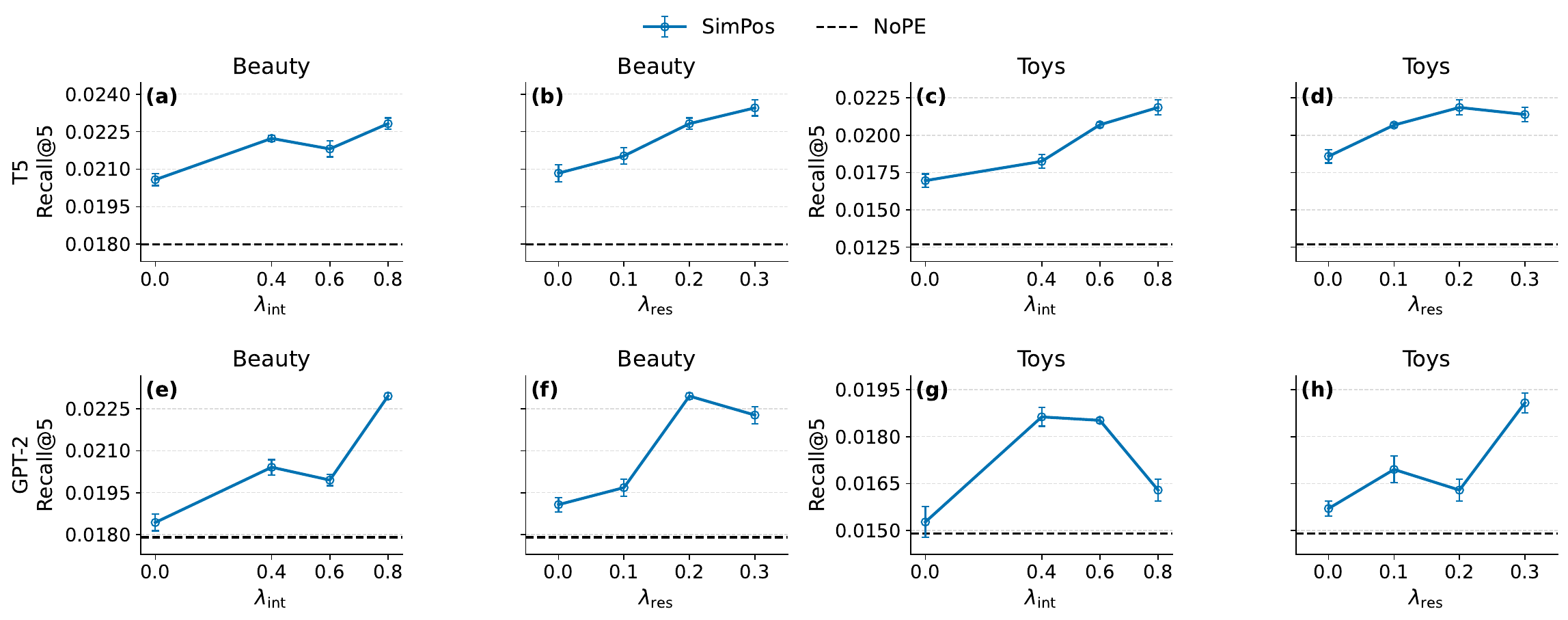}
  \caption{Effect of $\lambda_{\text{int}}$ and $\lambda_{\text{res}}$ on R@5 (mean $\pm$ std over five seeds) on Beauty and Toys with RQ-VAE. We vary $\lambda_{\text{int}} \in \{0, 0.4, 0.6, 0.8\}$ with $\lambda_{\text{res}} = 0.2$, and $\lambda_{\text{res}} \in \{0, 0.1, 0.2, 0.3\}$ with $\lambda_{\text{int}} = 0.8$. A coefficient of 0 removes that component, and NoPE sets both to 0.}
  \label{fig:slope_rqvae}
  \vspace{-6pt}
\end{figure*}

Table~\ref{tab:pe_comparison} reports the performance of different positional encodings under two backbones and two tokenizers.
We have the following observations.

\textbf{First, \mymodel{} consistently achieves the best performance across all settings.}
Across the 32 combinations of datasets, metrics, backbones, and tokenizers, \mymodel{} outperforms every baseline PE.
Compared with the strongest baseline in each group, \mymodel\ achieves relative gains of 5.00\% to 24.88\% in R@5 and 5.91\% to 24.64\% in N@5, with average gains of 12.66\% and 13.06\%, respectively.
Since the compared variants differ only in positional encoding, these improvements can be attributed to how positional information is modeled.
Moreover, \mymodel{} introduces no additional learnable parameters, so the gains do not stem from increased model capacity.

\textbf{Second, among existing PEs, the best choice varies across settings.}
None of the four baseline PEs outperforms the others across all settings, and their ranking changes with the dataset, backbone, and tokenizer.
Notably, the original PE of each backbone is often not the best baseline.
Under the T5 backbone, its default RelPE is the best baseline in only 6 of 16 cases.
Under the GPT-2 backbone, its default LPE is the best baseline in only 9 of 16 cases.
We attribute this to a shared limitation of these methods.
All of them define positions along the flattened token sequence, which conflates interaction order and residual order as discussed in Section~\ref{sec:rethinking_position}.
Therefore, replacing one flattened positional encoding with another yields limited and inconsistent changes.
In contrast, \mymodel\ models the two structures on separate positional axes and brings consistent improvements over all of them.

\textbf{Third, \mymodel{} generalizes across backbones and tokenizers.}
\mymodel{} improves both the encoder--decoder T5 and the decoder-only GPT-2, with average relative gains of 14.37\% and 11.35\%, respectively.
The gains are also comparable between RQ-VAE (12.73\%) and RQ-KMeans (12.98\%).
These results indicate that the benefit of \mymodel\ does not depend on a specific architecture or quantization method, since both tokenizers produce semantic IDs with the same interaction-level and residual-level structure.
We further evaluate \mymodel\ on the larger Amazon Books dataset, where it consistently outperforms all positional encoding baselines on both metrics (Appendix~\ref{app:books}).

\subsection{Further Analysis}
\label{sec:ablation}
We further analyze \mymodel\ from four aspects.
We first examine the contribution of each positional component (RQ2), then study the sensitivity to the bias strengths (RQ3), evaluate the robustness of \mymodel\ to different item embedding models (RQ4), and finally compare its two-axis relative bias with alternative positional designs (RQ5).

\subsubsection{Effect of Positional Components (RQ2)}

To examine the contribution of each component, we remove one bias at a time.
Setting $\lambda_{\text{int}} = 0$ keeps only the residual-level bias, and setting $\lambda_{\text{res}} = 0$ keeps only the interaction-level bias, which correspond to the leftmost points in Figure~\ref{fig:slope_rqvae} for RQ-VAE and Figure~\ref{fig:slope_rqkmeans} in Appendix~\ref{app:rqkmeans_sens} for RQ-KMeans.

\textbf{Each component is beneficial on its own.}
Both single-component variants outperform NoPE in all settings and with both tokenizers.
Notably, the residual-only variant improves over NoPE even though it carries no information about interaction order.
For example, on Toys with the T5 backbone, it raises R@5 from about 0.013 to 0.017 with RQ-VAE and from about 0.014 to 0.020 with RQ-KMeans.
This suggests that the residual structure within semantic IDs provides useful positional information on its own, even without interaction order.
Appendix~\ref{app:residual} further compares the linear residual-level bias with indicator, learnable, and reversed variants.
The results suggest that the benefit comes from encoding the distance and direction between residual levels, rather than from merely distinguishing them.

\textbf{The two components are complementary.}
The best results are always achieved when both coefficients are positive, and they exceed either single-component variant.
This is consistent with our analysis in Section~\ref{sec:rethinking_position}, since interaction order and residual order describe different structures of the flattened sequence and neither can replace the other.

\textbf{Positional modeling matters more for the encoder--decoder backbone.}
The gap between NoPE and \mymodel\ is considerably larger on T5 than on GPT-2.
For example, on Toys with RQ-VAE, NoPE reaches about 0.013 in R@5 on T5, compared with about 0.022 for the best \mymodel\ configuration, whereas the corresponding values on GPT-2 are about 0.015 and 0.019.
This difference arises because the bidirectional encoder of T5 is insensitive to input order without positional encoding, while the causal attention of GPT-2 implicitly captures part of the order information~\citep{haviv2022transformer}.
Even so, \mymodel\ still brings clear gains over NoPE on GPT-2, indicating that implicit order information is insufficient to capture the structure of SIDs.

\subsubsection{Sensitivity to Bias Strengths (RQ3)}
Figures~\ref{fig:slope_rqvae} and~\ref{fig:slope_rqkmeans} (Appendix~\ref{app:rqkmeans_sens}) also show how performance varies with the strength of each bias.

\textbf{The biases should guide attention rather than override it.}
With both tokenizers, performance generally improves as either coefficient increases from zero and peaks at moderate values, with $\lambda_{\text{int}} \in [0.4, 0.8]$ and $\lambda_{\text{res}} \in [0.1, 0.3]$.
In some settings, an overly large coefficient reduces performance, as a strong bias tends to suppress attention to distant items or residual levels.
For example, with RQ-KMeans on Beauty, R@5 on T5 decreases when $\lambda_{\text{res}}$ increases from 0.2 to 0.3.
Nevertheless, all configurations with positive coefficients outperform NoPE in every setting.

\textbf{A strong recency preference is beneficial.}
The preferred values of $\lambda_{\text{int}}$ are relatively large.
With $\lambda_{\text{int}} = 0.8$, the attention logit to a code five items earlier is reduced by 4, so attention concentrates on the most recent interactions.
The effectiveness of such a strong bias suggests that next-item prediction on these datasets relies mainly on recent interactions, which \mymodel\ encodes explicitly.
However, recency alone does not explain the gains, as a tuned single-axis bias remains less effective than \mymodel\ (\S\ref{sec:separate_axes}).

\textbf{The two coefficients operate at different scales.}
$\lambda_{\text{res}}$ has a smaller effective range than $\lambda_{\text{int}}$, since residual distances are bounded by the number of codes per item, whereas interaction distances grow with sequence length.

\subsubsection{Effect of Embedding Models (RQ4)}

\begin{table*}[t]
  \centering

  \caption{Performance of different positional encodings (PE) with the T5 backbone, using \texttt{sentence-T5-XL}~\citep{ni2022sentence} as the item embedding model. The backbone architecture and training hyperparameters are the same as in Table~\ref{tab:pe_comparison}. In each group, the best result is in \textbf{bold}, and the second best is \underline{underlined}. \textit{Improv.} is the relative gain of \mymodel\ over the best baseline.}
 \vspace{-2pt}
  \label{tab:pe_comparison_tiger_sentence_t5}

  \setlength{\tabcolsep}{1.5mm}

  \resizebox{\textwidth}{!}{%
  \begin{tabular}{@{}llcccccccc@{}}
    \toprule

    \multirow{2}{*}{\textbf{Tokenizer}} &
    \multirow{2}{*}{\textbf{PE}} &
    \multicolumn{2}{c}{\textbf{Beauty}} &
    \multicolumn{2}{c}{\textbf{Toys}} &
    \multicolumn{2}{c}{\textbf{Sports}} &
    \multicolumn{2}{c}{\textbf{CDs}} \\

    \cmidrule(lr){3-4}
    \cmidrule(lr){5-6}
    \cmidrule(lr){7-8}
    \cmidrule(l){9-10}

    & & R@5 & N@5 & R@5 & N@5 & R@5 & N@5 & R@5 & N@5 \\

    \midrule

    \multicolumn{10}{@{}c}{%
      \textit{T5 backbone (Original positional encoding is RelPE)}} \\

    \midrule

    \multirow{6}{*}{RQ-VAE}
      & LPE
      & 0.0211 & 0.0132
      & 0.0194 & 0.0125
      & 0.0201 & 0.0131
      & 0.0278 & 0.0175 \\

      & RelPE
      & \underline{0.0217} & \underline{0.0137}
      & \underline{0.0200} & \underline{0.0128}
      & 0.0208 & 0.0135
      & \underline{0.0291} & 0.0182 \\

      & RoPE
      & 0.0204 & 0.0129
      & 0.0189 & 0.0120
      & 0.0195 & 0.0127
      & 0.0274 & 0.0171 \\

      & ALiBi
      & 0.0213 & 0.0130
      & 0.0196 & 0.0122
      & \underline{0.0213} & \underline{0.0137}
      & 0.0282 & \underline{0.0187} \\

      & \textbf{SimPos}
      & \textbf{0.0234} & \textbf{0.0152}
      & \textbf{0.0212} & \textbf{0.0138}
      & \textbf{0.0236} & \textbf{0.0156}
      & \textbf{0.0351} & \textbf{0.0225} \\

    \rowcolor{gray!12}
    \cellcolor{white}
      & \textit{Improv.}
      & \textit{+7.83\%}  & \textit{+10.95\%}
      & \textit{+6.00\%}  & \textit{+7.81\%}
      & \textit{+10.80\%} & \textit{+13.87\%}
      & \textit{+20.62\%} & \textit{+20.32\%} \\

    \cmidrule{1-10}

    \multirow{6}{*}{RQ-KMeans}
      & LPE
      & 0.0208 & 0.0131
      & 0.0197 & 0.0126
      & 0.0199 & 0.0133
      & 0.0326 & 0.0203 \\

      & RelPE
      & 0.0215 & 0.0135
      & \underline{0.0203} & \underline{0.0130}
      & 0.0207 & 0.0137
      & \underline{0.0342} & \underline{0.0210} \\

      & RoPE
      & 0.0202 & 0.0127
      & 0.0191 & 0.0122
      & 0.0194 & 0.0129
      & 0.0321 & 0.0197 \\

      & ALiBi
      & \underline{0.0220} & \underline{0.0139}
      & 0.0186 & 0.0119
      & \underline{0.0214} & \underline{0.0141}
      & 0.0332 & 0.0208 \\

      & \textbf{SimPos}
      & \textbf{0.0242} & \textbf{0.0151}
      & \textbf{0.0231} & \textbf{0.0150}
      & \textbf{0.0252} & \textbf{0.0166}
      & \textbf{0.0363} & \textbf{0.0228} \\

    \rowcolor{gray!12}
    \cellcolor{white}
      & \textit{Improv.}
      & \textit{+10.00\%} & \textit{+8.63\%}
      & \textit{+13.79\%} & \textit{+15.38\%}
      & \textit{+17.76\%} & \textit{+17.73\%}
      & \textit{+6.14\%}  & \textit{+8.57\%} \\

    \bottomrule
  \end{tabular}%
  }
  \vspace{-6pt}
\end{table*}

To examine whether the gains of \mymodel\ depend on a specific embedding model, we replace \texttt{Qwen3-Embedding-0.6B\textsuperscript{\ref{fn:qwen}}} with \texttt{sentence-T5-XL}\footnote{https://huggingface.co/sentence-transformers/sentence-t5-xl}, the item encoder adopted by TIGER~\citep{tiger}, and keep the backbone architecture and training hyperparameters unchanged.
Table~\ref{tab:pe_comparison_tiger_sentence_t5} reports the results with the T5 backbone, which can be compared with the T5 results in Table~\ref{tab:pe_comparison}.

\textbf{\mymodel\ is robust to the choice of embedding model.}
It remains the best positional encoding, with average relative gains of 12.28\% with RQ-VAE and 12.25\% with RQ-KMeans, which are close to those obtained with \texttt{Qwen3-Embedding-0.6B} (13.48\% and 15.25\%, respectively).

\textbf{Embedding quality and positional modeling are complementary.}
Neither embedding model is consistently better across datasets.
With RQ-KMeans, for example, \texttt{Qwen3-Embedding-0.6B} yields higher R@5 on Beauty (0.0271 vs.\ 0.0242), whereas \texttt{sentence-T5-XL} yields higher R@5 on CDs (0.0363 vs.\ 0.0331), and the best baseline PE also changes with the embedding model.
In contrast, the gains of \mymodel\ remain stable.
The baselines differ only in how they parameterize the same flattened axis, so their small differences shift with the semantic content of the codes.
\mymodel\ instead exploits the interaction-level and residual-level structure determined by the tokenizer, and its advantage far exceeds the differences among baselines.
The two choices can therefore be made independently, since a stronger embedding model improves all methods and \mymodel\ adds a consistent gain on top.

\subsubsection{Role of the Two-Axis Relative Bias (RQ5)}
\label{sec:separate_axes}

To assess whether \mymodel's gains arise merely from access to the two coordinates $(t,k)$ or a better-tuned locality scale, we consider three baselines:
(1) factorized absolute PE, which adds $E_{\mathrm{item}}(t)+E_{\mathrm{level}}(k)$ to the input;
(2) tuned ALiBi, which scales the default head-specific slopes $m_h$ by a factor $\alpha$, giving $b_h(q,k)=-\alpha\,m_h\,|p_q-p_k|$; and
(3) a flattened linear relative bias, $b(q,k)=-\lambda\,|p_q-p_k|$, with a single slope shared across heads, defined over the flattened token position $p$.
We sweep $\alpha$ and $\lambda$ to tune the strength of the two single-axis biases (Appendix~\ref{app:baselines}).

As shown in Table~\ref{tab:app_baselines}, factorized absolute PE outperforms LPE in only 4 of 32 setting--metric combinations and ties in 5, while \mymodel\ outperforms it by $8.19\%$--$38.40\%$ across all settings.
Thus, providing interaction order and residual level as additive input embeddings does not yield the same benefits as incorporating their structure through relative attention biases.
Moreover, tuning either single-axis bias yields a median improvement of less than $2\%$ over default ALiBi, and \mymodel\ still outperforms the best tuned single-axis bias by $2.70\%$--$30.53\%$.
Together, these results suggest that the advantage of \mymodel\ stems from its two-axis relative bias itself, rather than from access to the two coordinates or the strength of the locality bias.

\vspace{-6pt}
\section{Conclusion}
In this paper, we study positional encoding in semantic ID-based generative recommendation.
We show that serializing multi-code items into a flattened sequence places interaction order and residual order on a single token-level positional axis, where conventional positional encodings do not explicitly model their distinct structural roles.
To address this, we propose \mymodel, a plug-and-play positional modeling method that encodes interaction-level and residual-level distances as two separate additive attention biases, without additional learnable parameters.
Experiments on four real-world datasets with two Transformer backbones and two residual tokenizers show that \mymodel\ consistently outperforms existing positional encoding methods.
These findings suggest that positional encoding deserves explicit design in semantic ID-based generative recommendation rather than being directly inherited from the backbone.

\textbf{Limitations and Future Work.}
\mymodel\ is designed for residual semantic IDs, whose codes follow a coarse-to-fine order.
A natural next step is to extend structure-aware positional modeling to non-hierarchical semantic IDs, such as product-quantized IDs~\citep{rpg}.
In addition, the two bias coefficients are fixed hyperparameters shared across attention heads and layers.
Learning them or making them head-specific may further improve flexibility.
Finally, industrial-scale evaluation remains future work.

\bibliography{reference}
\bibliographystyle{icml2027}

\newpage
\appendix
\onecolumn
\section{Additional Experimental Details}
\label{app:setup}

\subsection{Datasets}
\label{app:datasets}
The four datasets in \S\ref{sec:experiment} correspond to the ``Beauty and Personal Care'', ``Toys and Games'', ``Sports and Outdoors'', and ``CDs and Vinyl'' categories of Amazon Reviews 2023, and the larger Books dataset used in Appendix~\ref{app:books} corresponds to the ``Books'' category.
Following prior work~\citep{tiger,rpg}, we treat each review as an interaction and sort each user's interactions chronologically.
We remove duplicate user--item interactions and apply iterative $k$-core filtering to ensure sufficient behavioral signals.
Table~\ref{tab:datasets} summarizes the statistics of the processed datasets.

\begin{table}[ht]
  \centering
  \caption{Statistics of the processed datasets.}
  \vspace{-4pt}
  \label{tab:datasets}
  \small
  \setlength{\tabcolsep}{4pt}
  \begin{tabular}{lcccccc}
    \toprule
    Datasets
    & \makecell{$k$-core}
    & \makecell{\# Users}
    & \makecell{\# Items}
    & \makecell{\# Interactions}
    & \makecell{Density}
    & \makecell{Average \\ Sequence Length} \\
    \midrule
    Beauty & 15 & 8,425   & 11,509  & 304,396   & 0.314\% & 36.13 \\
    Toys   & 10 & 41,744  & 31,132  & 728,215   & 0.056\% & 17.44 \\
    Sports & 10 & 17,906  & 11,247  & 267,830   & 0.133\% & 14.96 \\
    CDs    & 10 & 22,170  & 20,089  & 519,283   & 0.117\% & 23.42 \\
    Books  & 10 & 139,089 & 113,175 & 3,344,074 & 0.021\% & 24.04 \\
    \bottomrule
  \end{tabular}
\end{table}

\subsection{Selected Coefficients}
\label{app:coefficients}
Table~\ref{tab:best_hyperparameters} lists the coefficients $(\lambda_{\text{int}}, \lambda_{\text{res}})$ of \mymodel\ selected on validation N@5 for each setting in Table~\ref{tab:pe_comparison}.

\begin{table}[ht]
\centering
\small
\caption{Selected coefficients $(\lambda_{\mathrm{int}}, \lambda_{\mathrm{res}})$ of \mymodel\ for each setting in Table~\ref{tab:pe_comparison}.}
\vspace{-4pt}
\label{tab:best_hyperparameters}
\setlength{\tabcolsep}{5pt}
\begin{tabular}{@{}llcccccccc@{}}
\toprule
\multirow{2.5}{*}{\textbf{Backbone}} & \multirow{2.5}{*}{\textbf{Tokenizer}} & \multicolumn{2}{c}{\textbf{Beauty}} & \multicolumn{2}{c}{\textbf{Toys}} & \multicolumn{2}{c}{\textbf{Sports}} & \multicolumn{2}{c}{\textbf{CDs}} \\
\cmidrule(l){3-4} \cmidrule(l){5-6} \cmidrule(l){7-8} \cmidrule(l){9-10}
& & $\lambda_{\text{int}}$ & $\lambda_{\text{res}}$ & $\lambda_{\text{int}}$ & $\lambda_{\text{res}}$ & $\lambda_{\text{int}}$ & $\lambda_{\text{res}}$ & $\lambda_{\text{int}}$ & $\lambda_{\text{res}}$ \\
\midrule
\multirow{2}{*}{T5}    & RQ-VAE    & 0.8 & 0.2 & 0.8 & 0.1 & 0.6 & 0.2 & 0.6 & 0.2 \\
                       & RQ-KMeans & 0.6 & 0.1 & 0.8 & 0.1 & 0.8 & 0.1 & 0.6 & 0.1 \\
\midrule
\multirow{2}{*}{GPT-2} & RQ-VAE    & 0.8 & 0.2 & 0.6 & 0.2 & 0.6 & 0.1 & 0.6 & 0.1 \\
                       & RQ-KMeans & 0.6 & 0.1 & 0.6 & 0.2 & 0.6 & 0.2 & 0.6 & 0.1 \\
\bottomrule
\end{tabular}
\end{table}

\subsection{Component Ablation and Bias Sensitivity with RQ-KMeans}
\label{app:rqkmeans_sens}
\begin{figure}[H]
  \centering
  \includegraphics[width=0.9\textwidth]{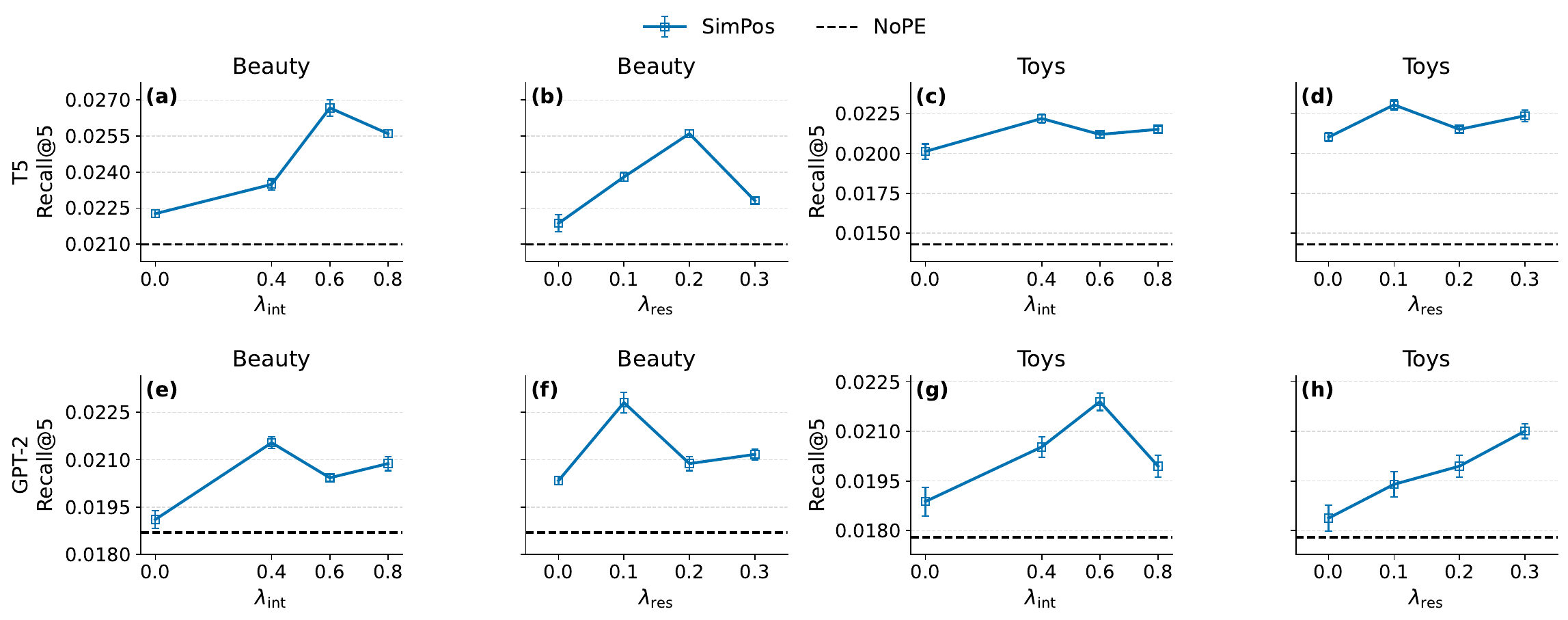}
  \caption{Effect of $\lambda_{\text{int}}$ and $\lambda_{\text{res}}$ on R@5 on Beauty and Toys with RQ-KMeans. The setup is the same as in Figure~\ref{fig:slope_rqvae}.}
  \label{fig:slope_rqkmeans}
\end{figure}

Figure~\ref{fig:slope_rqkmeans} reports the effect of $\lambda_{\text{int}}$ and $\lambda_{\text{res}}$ with the RQ-KMeans tokenizer, complementing the RQ-VAE results in Figure~\ref{fig:slope_rqvae}.
The trends are consistent with those discussed in \S\ref{sec:ablation}. Either bias alone already improves over NoPE, indicating that each axis carries useful positional information. Furthermore, the two relative biases complement each other, yielding the best results only when both are active. However, performance drops once a coefficient becomes too large.

\section{Additional Experiments on Amazon Books}
\label{app:books}

To further evaluate \mymodel\ on a larger-scale dataset, we conduct additional experiments on the ``Books'' category of Amazon Reviews 2023, whose statistics are given in Table~\ref{tab:datasets}.
We follow the experimental setup in \S\ref{sec:experiment_setup}, including data preprocessing, backbones (T5 and GPT-2), tokenizers (RQ-VAE and RQ-KMeans), item embedding model (\texttt{Qwen3-Embedding-0.6B}), and training configuration. We set $\lambda_{\text{int}}=0.6$ and $\lambda_{\text{res}}=0.1$ for \mymodel\ on Books.
As shown in Table~\ref{tab:books}, \mymodel\ achieves the best performance across all settings, with relative gains over the strongest baseline of 8.41\%--16.32\% in R@5 and 6.47\%--17.36\% in N@5, consistent with the results in Table~\ref{tab:pe_comparison}.

\begin{table}[h]
  \centering
  \caption{Performance comparison on Amazon Books with two backbones and two tokenizers. Best in \textbf{bold}, second best \underline{underlined}. \textit{Improv.} is the relative gain of \mymodel\ over the best baseline. Results are means over five training seeds (std $\leq 0.0007$). \mymodel\ achieves statistically significant improvements under a paired t-test (Holm-adjusted $p < 0.001$).}
  \vspace{-2pt}
  \label{tab:books}
  \setlength{\tabcolsep}{1.5mm}
  \small
  \begin{tabular}{@{}lcccccccc@{}}
    \toprule
    \multirow{3.5}{*}{\textbf{PE}} & \multicolumn{4}{c}{\textit{T5 backbone}} & \multicolumn{4}{c}{\textit{GPT-2 backbone}} \\
    \cmidrule(l){2-5} \cmidrule(l){6-9}
    & \multicolumn{2}{c}{RQ-VAE} & \multicolumn{2}{c}{RQ-KMeans} & \multicolumn{2}{c}{RQ-VAE} & \multicolumn{2}{c}{RQ-KMeans} \\
    \cmidrule(l){2-3} \cmidrule(l){4-5} \cmidrule(l){6-7} \cmidrule(l){8-9}
    & R@5 & N@5 & R@5 & N@5 & R@5 & N@5 & R@5 & N@5 \\
    \midrule\midrule
    LPE   & 0.0308 & 0.0192 & 0.0312 & 0.0198 & 0.0184 & 0.0114 & 0.0210 & 0.0132 \\
    RelPE & 0.0381 & \underline{0.0253} & \underline{0.0393} & \underline{0.0242} & \underline{0.0190} & 0.0118 & 0.0199 & 0.0128 \\
    RoPE  & 0.0342 & 0.0231 & 0.0339 & 0.0218 & 0.0175 & 0.0107 & 0.0213 & 0.0136 \\
    ALiBi & \underline{0.0397} & 0.0247 & 0.0335 & 0.0212 & 0.0186 & \underline{0.0122} & \underline{0.0214} & \underline{0.0139} \\
    \textbf{SimPos} & \textbf{0.0441} & \textbf{0.0289} & \textbf{0.0431} & \textbf{0.0284} & \textbf{0.0221} & \textbf{0.0138} & \textbf{0.0232} & \textbf{0.0148} \\
    \rowcolor{gray!12}
    \textit{Improv.} & \textit{+11.08\%} & \textit{+14.23\%} & \textit{+9.67\%} & \textit{+17.36\%} & \textit{+16.32\%} & \textit{+13.11\%} & \textit{+8.41\%} & \textit{+6.47\%} \\
    \bottomrule
  \end{tabular}
  \vspace{-4pt}
\end{table}

\section{Ablation on Residual-Level Bias Formulations}
\label{app:residual}

The residual-level bias of \mymodel\ decays linearly with the level distance $|k-j|$ (Eq.~(\ref{eq:simpos_residual})), motivated by the sequential structure of residual quantization.
To assess whether this formulation matters beyond simply distinguishing different levels, we consider three alternatives while keeping the interaction-level bias and all other settings unchanged:
(1) \textbf{Indicator}, which assigns zero bias to tokens at the same residual level and $-\lambda_{\mathrm{res}}$ otherwise;
(2) \textbf{Learnable}, which replaces the fixed distance-based bias with a learnable level-pair bias matrix; and
(3) \textbf{Reversed}, which reverses the distance-based bias by assigning larger penalties to closer residual levels.

Table~\ref{tab:app_residual} compares these formulations on Beauty and Toys across both backbones and tokenizers, using the same five training seeds as Table~\ref{tab:pe_comparison}. For Indicator, we select $\lambda_{\mathrm{res}}\in\{0.1,0.2,0.3\}$ based on validation performance for each setting. Reversed uses the corresponding $\lambda_{\mathrm{res}}$ from \mymodel.
\mymodel\ achieves the best performance in all 16 setting--metric combinations. Indicator generally outperforms Learnable and Reversed but remains below \mymodel. These results suggest that explicitly modeling distances between residual levels is more effective than merely distinguishing levels or learning unconstrained level-pair biases, while the poorer performance of Reversed further highlights the importance of the bias direction.

\begin{table}[ht]
  \centering
  \caption{Ablation on the formulation of the residual-level bias on Beauty and Toys. All variants share the interaction-level bias of \mymodel\ and differ only in the residual-level term. Results are means over five training seeds (std $\leq 0.0004$). The best results are highlighted in \textbf{bold}. \mymodel\ significantly outperforms all variants under a paired t-test (all Holm-adjusted $p < 0.01$).}
  \label{tab:app_residual}
  \setlength{\tabcolsep}{1.1mm}
  \resizebox{\textwidth}{!}{
  \begin{tabular}{@{}lcccccccccccccccc@{}}
    \toprule
    \multirow{3.5}{*}{\textbf{Method}} & \multicolumn{4}{c}{\textbf{T5 / RQ-VAE}} & \multicolumn{4}{c}{\textbf{T5 / RQ-KMeans}} & \multicolumn{4}{c}{\textbf{GPT-2 / RQ-VAE}} & \multicolumn{4}{c}{\textbf{GPT-2 / RQ-KMeans}} \\
    \cmidrule(l){2-5} \cmidrule(l){6-9} \cmidrule(l){10-13} \cmidrule(l){14-17}
    & \multicolumn{2}{c}{Beauty} & \multicolumn{2}{c}{Toys} & \multicolumn{2}{c}{Beauty} & \multicolumn{2}{c}{Toys} & \multicolumn{2}{c}{Beauty} & \multicolumn{2}{c}{Toys} & \multicolumn{2}{c}{Beauty} & \multicolumn{2}{c}{Toys} \\
    \cmidrule(l){2-3} \cmidrule(l){4-5} \cmidrule(l){6-7} \cmidrule(l){8-9} \cmidrule(l){10-11} \cmidrule(l){12-13} \cmidrule(l){14-15} \cmidrule(l){16-17}
    & R@5 & N@5 & R@5 & N@5 & R@5 & N@5 & R@5 & N@5 & R@5 & N@5 & R@5 & N@5 & R@5 & N@5 & R@5 & N@5 \\
    \midrule
Indicator
& 0.0217 & 0.0144 & 0.0198 & 0.0128
& 0.0259 & 0.0164 & 0.0217 & 0.0151
& 0.0221 & 0.0143 & 0.0178 & 0.0116
& 0.0219 & 0.0145 & 0.0208 & 0.0133 \\
    Learnable & 0.0212 & 0.0142 & 0.0195 & 0.0126 & 0.0252 & 0.0159 & 0.0217 & 0.0148 & 0.0215 & 0.0140 & 0.0174 & 0.0113 & 0.0213 & 0.0141 & 0.0202 & 0.0129 \\
    Reversed & 0.0204 & 0.0136 & 0.0187 & 0.0121 & 0.0243 & 0.0154 & 0.0208 & 0.0143 & 0.0208 & 0.0134 & 0.0166 & 0.0109 & 0.0206 & 0.0137 & 0.0195 & 0.0125 \\
    \midrule
    \mymodel & \textbf{0.0228} & \textbf{0.0152} & \textbf{0.0207} & \textbf{0.0135} & \textbf{0.0271} & \textbf{0.0173} & \textbf{0.0231} & \textbf{0.0159} & \textbf{0.0230} & \textbf{0.0150} & \textbf{0.0185} & \textbf{0.0120} & \textbf{0.0227} & \textbf{0.0152} & \textbf{0.0219} & \textbf{0.0139} \\
    \bottomrule
  \end{tabular}}
\end{table}

\section{Additional Positional Encoding Baselines}
\label{app:baselines}

We provide additional experimental details and full results for the three baselines introduced in \S\ref{sec:separate_axes}.
All three replace the backbone's original positional encoding (the relative bias of T5 and the learned absolute embedding of GPT-2), and otherwise use exactly the same semantic IDs, backbones, training configuration, and five training seeds as Table~\ref{tab:pe_comparison}.
For tuned ALiBi, we search the global scaling factor $\alpha\in\{0.2,0.4,0.6,0.8,1,2,4\}$. For the flattened linear relative bias (FlatBias), we search $\lambda\in\{0.2, 0.4, 0.6, 0.8, 1.0, 1.5, 2.0\}$, a tuning budget comparable to the nine $(\lambda_{\mathrm{int}},\lambda_{\mathrm{res}})$ configurations of \mymodel. Factorized absolute PE introduces no additional hyperparameters to tune.

\begin{figure}[ht]
  \vspace{-10pt}
  \centering
  \includegraphics[width=0.8\textwidth]{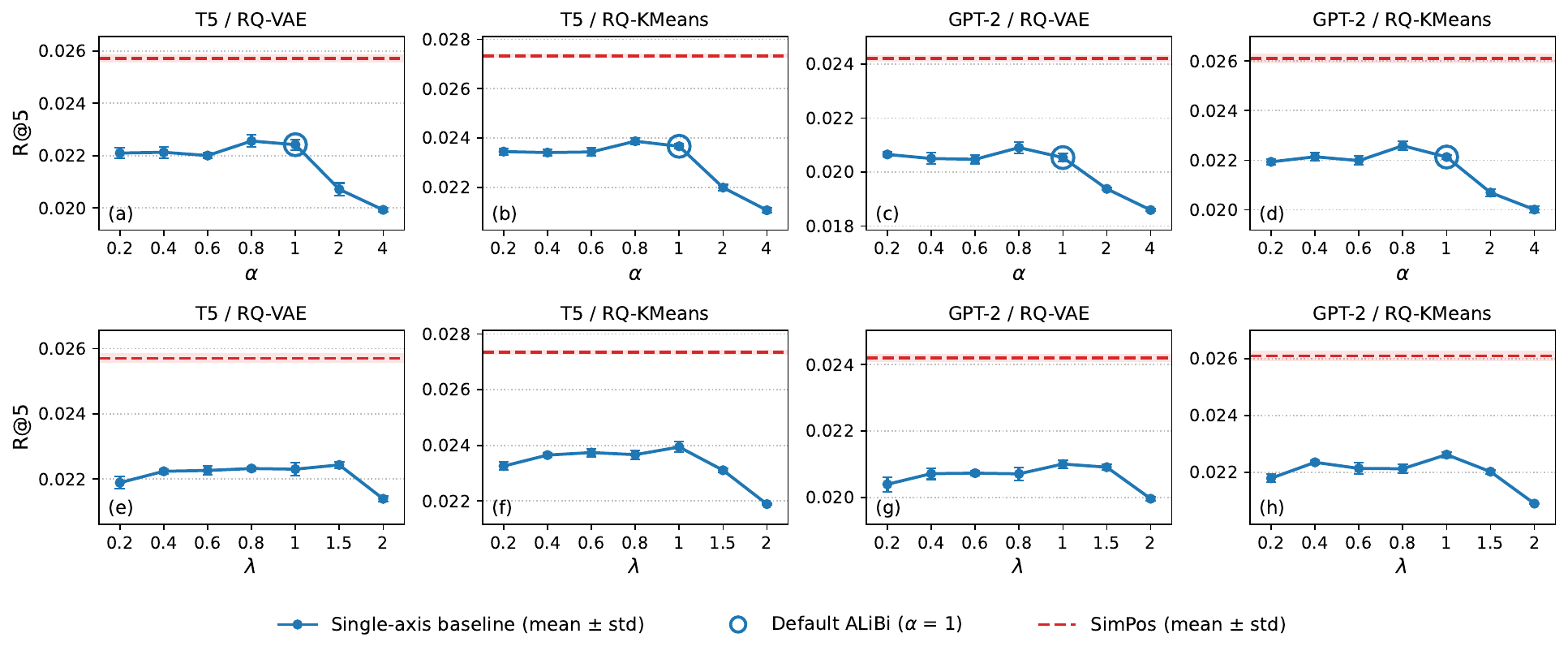}
  \vspace{-8pt}
  \caption{Sensitivity of tuned ALiBi (top, $\alpha$) and FlatBias (bottom, $\lambda$) to bias strength across backbone--tokenizer settings, averaged over four datasets (mean $\pm$ std over five seeds). The hollow marker at $\alpha=1$ denotes default ALiBi.}
  \label{fig:app_sweeps}
\end{figure}

We further analyze the sensitivity of tuned ALiBi and FlatBias to their respective bias-strength coefficients. As shown in Figure~\ref{fig:app_sweeps}, both baselines exhibit variations in performance across different coefficient values but remain below \mymodel\ throughout the evaluated ranges. This suggests that their performance gap with \mymodel\ cannot be explained solely by insufficient tuning of the bias strength.

\begin{table}[H]
  \vspace{-12pt}
  \centering
  \caption{Per-dataset comparison of \mymodel\ with additional positional encoding baselines. Tuned ALiBi and FlatBias are reported at their best $\alpha$ and $\lambda$ for each setting. Best in \textbf{bold}, second best \underline{underlined}. \textit{Improv.} is the relative gain of \mymodel\ over the best baseline. Results are means over five training seeds (std $\leq 0.0006$). \mymodel\ achieves statistically significant improvements under a paired t-test (Holm-adjusted $p < 0.01$).}
  \label{tab:app_baselines}
  \vspace{-6pt}
  \footnotesize
  \renewcommand{\arraystretch}{0.85}
  \setlength{\tabcolsep}{1.6mm}
  \begin{tabular}{@{}llcccccccc@{}}
    \toprule
    \multirow{2.5}{*}{\textbf{Tokenizer}} & \multirow{2.5}{*}{\textbf{PE}} & \multicolumn{2}{c}{\textbf{Beauty}} & \multicolumn{2}{c}{\textbf{Toys}} & \multicolumn{2}{c}{\textbf{Sports}} & \multicolumn{2}{c}{\textbf{CDs}} \\
    \cmidrule(l){3-4} \cmidrule(l){5-6} \cmidrule(l){7-8} \cmidrule(l){9-10}
    & & R@5 & N@5 & R@5 & N@5 & R@5 & N@5 & R@5 & N@5 \\
    \midrule\midrule
    \multicolumn{10}{@{}c}{\textit{T5 backbone}} \\
    \midrule
    \multirow{7}{*}{RQ-VAE}
      & LPE & 0.0202 & 0.0129 & 0.0182 & 0.0109 & 0.0183 & \underline{0.0122} & 0.0301 & 0.0183 \\
      & Factorized PE & 0.0199 & 0.0129 & 0.0181 & 0.0109 & 0.0180 & 0.0120 & 0.0297 & 0.0178 \\
      & ALiBi (default) & 0.0213 & 0.0136 & 0.0186 & 0.0111 & \underline{0.0190} & 0.0117 & \underline{0.0308} & 0.0190 \\
      & Tuned ALiBi & 0.0220 & 0.0138 & \underline{0.0189} & 0.0111 & \underline{0.0190} & 0.0120 & \underline{0.0308} & 0.0190 \\
      & FlatBias & \underline{0.0222} & \underline{0.0140} & 0.0188 & \underline{0.0112} & 0.0189 & 0.0119 & 0.0302 & \underline{0.0193} \\
      & \textbf{\mymodel} & \textbf{0.0228} & \textbf{0.0152} & \textbf{0.0207} & \textbf{0.0135} & \textbf{0.0248} & \textbf{0.0154} & \textbf{0.0345} & \textbf{0.0214} \\
    \rowcolor{gray!12}
    \cellcolor{white}
      & \textit{Improv.} & \textit{+2.70\%} & \textit{+8.57\%} & \textit{+9.52\%} & \textit{+20.54\%} & \textit{+30.53\%} & \textit{+26.23\%} & \textit{+12.01\%} & \textit{+10.88\%} \\
    \cmidrule{1-10}
    \multirow{7}{*}{RQ-KMeans}
      & LPE & 0.0203 & 0.0126 & 0.0209 & 0.0133 & 0.0200 & 0.0130 & 0.0294 & 0.0182 \\
      & Factorized PE & 0.0202 & 0.0125 & 0.0208 & 0.0130 & 0.0195 & 0.0127 & 0.0290 & 0.0178 \\
      & ALiBi (default) & 0.0217 & 0.0139 & 0.0211 & \underline{0.0143} & \underline{0.0217} & 0.0138 & 0.0301 & 0.0186 \\
      & Tuned ALiBi & \underline{0.0223} & 0.0139 & 0.0214 & \underline{0.0143} & \underline{0.0217} & \underline{0.0141} & \underline{0.0302} & \underline{0.0194} \\
      & FlatBias & 0.0221 & \underline{0.0140} & \underline{0.0217} & \underline{0.0143} & \underline{0.0217} & 0.0140 & \underline{0.0302} & \underline{0.0194} \\
      & \textbf{\mymodel} & \textbf{0.0271} & \textbf{0.0173} & \textbf{0.0231} & \textbf{0.0159} & \textbf{0.0261} & \textbf{0.0172} & \textbf{0.0331} & \textbf{0.0215} \\
    \rowcolor{gray!12}
    \cellcolor{white}
      & \textit{Improv.} & \textit{+21.52\%} & \textit{+23.57\%} & \textit{+6.45\%} & \textit{+11.19\%} & \textit{+20.28\%} & \textit{+21.99\%} & \textit{+9.60\%} & \textit{+10.82\%} \\
    \midrule
    \multicolumn{10}{@{}c}{\textit{GPT-2 backbone}} \\
    \midrule
    \multirow{7}{*}{RQ-VAE}
      & LPE & 0.0197 & 0.0128 & 0.0172 & \underline{0.0102} & 0.0213 & \underline{0.0137} & \underline{0.0285} & \underline{0.0177} \\
      & Factorized PE & \underline{0.0198} & 0.0125 & 0.0171 & \underline{0.0102} & 0.0211 & 0.0134 & 0.0276 & 0.0174 \\
      & ALiBi (default) & 0.0185 & 0.0130 & 0.0165 & 0.0098 & 0.0208 & 0.0127 & 0.0264 & 0.0167 \\
      & Tuned ALiBi & 0.0187 & 0.0132 & 0.0173 & \underline{0.0102} & 0.0211 & 0.0127 & 0.0266 & 0.0170 \\
      & FlatBias & 0.0187 & \underline{0.0133} & \underline{0.0174} & 0.0100 & \underline{0.0215} & 0.0129 & 0.0267 & 0.0172 \\
      & \textbf{\mymodel} & \textbf{0.0230} & \textbf{0.0150} & \textbf{0.0185} & \textbf{0.0120} & \textbf{0.0234} & \textbf{0.0156} & \textbf{0.0319} & \textbf{0.0203} \\
    \rowcolor{gray!12}
    \cellcolor{white}
      & \textit{Improv.} & \textit{+16.16\%} & \textit{+12.78\%} & \textit{+6.32\%} & \textit{+17.65\%} & \textit{+8.84\%} & \textit{+13.87\%} & \textit{+11.93\%} & \textit{+14.69\%} \\
    \cmidrule{1-10}
    \multirow{7}{*}{RQ-KMeans}
      & LPE & 0.0208 & 0.0133 & 0.0195 & 0.0125 & 0.0214 & \underline{0.0146} & \underline{0.0303} & 0.0191 \\
      & Factorized PE & 0.0208 & 0.0133 & \underline{0.0196} & \underline{0.0127} & \underline{0.0215} & 0.0144 & 0.0300 & 0.0187 \\
      & ALiBi (default) & 0.0208 & 0.0131 & 0.0190 & 0.0122 & 0.0200 & 0.0135 & 0.0288 & 0.0195 \\
      & Tuned ALiBi & \underline{0.0212} & 0.0134 & 0.0193 & 0.0122 & 0.0208 & 0.0140 & 0.0291 & 0.0199 \\
      & FlatBias & 0.0210 & \underline{0.0136} & 0.0193 & 0.0121 & 0.0207 & 0.0140 & 0.0295 & \underline{0.0201} \\
      & \textbf{\mymodel} & \textbf{0.0227} & \textbf{0.0152} & \textbf{0.0219} & \textbf{0.0139} & \textbf{0.0254} & \textbf{0.0169} & \textbf{0.0343} & \textbf{0.0215} \\
    \rowcolor{gray!12}
    \cellcolor{white}
      & \textit{Improv.} & \textit{+7.08\%} & \textit{+11.76\%} & \textit{+11.73\%} & \textit{+9.45\%} & \textit{+18.14\%} & \textit{+15.75\%} & \textit{+13.20\%} & \textit{+6.97\%} \\
    \bottomrule
  \end{tabular}
\end{table}

\end{document}